\documentclass{emulateapj}
\usepackage{times}
\usepackage{natbib}
\usepackage{apjfonts}
\usepackage{amssymb}
\usepackage{epsfig}

\newcommand{\sgra}{Sgr~A*}

\newcommand{\chandra}{{\em Chandra}}
\defcitealias{kvk05}{KvK05a}
\defcitealias{kvk05b}{KvK05b}
\defcitealias{vkk08}{vKK08}
\defcitealias{zct+05}{Z+05}

\begin{document}

\shorttitle{A 15~GHz Search for Pulsars Within the Central Parsec of Sgr~A*}
\shortauthors{Macquart et al.}

\title{A High-Frequency Search for Pulsars Within the Central Parsec of Sgr~A*}

\author{J-P. Macquart\altaffilmark{1}, N. Kanekar\altaffilmark{2}, 
D. A. Frail\altaffilmark{3}, and S. M. Ransom\altaffilmark{4}}

\altaffiltext{1}{NRAO Jansky Fellow, Department of Astronomy,
  California Institute of Technology, Pasadena, CA 91125, USA; ICRAR/Curtin
  University of Technology, Bentley, WA 6845, Australia; J.Macquart@curtin.edu.au}

\altaffiltext{2}{NRAO Jansky Fellow, National Radio Astronomy Observatory, 1003 Lopezville
  Road, Socorro, NM 87801, USA; Ramanujan Fellow, National Centre for Radio Astrophysics, 
Tata Institute of Fundamental Research, Ganeshkhind, Pune -- 411007, India}

\altaffiltext{3}{National Radio Astronomy Observatory, 1003 Lopezville
  Road, Socorro, NM 87801, USA}

\altaffiltext{4}{National Radio Astronomy Observatory, 520 Edgemont Road, 
Charlottesville, VA 29903, USA}

\begin{abstract}
  We report results from a deep high-frequency search for pulsars
  within the central parsec of Sgr~A* using the Green Bank
    Telescope.  The observing frequency of $15$\,GHz was chosen to
  maximize the likelihood of detecting normal pulsars (i.e. with
  periods of $\sim 500$\,ms and spectral indices of $\sim -1.7$) close
  to Sgr~A*, that might be used as probes of gravity in the
  strong-field regime; this is the highest frequency used for such
  pulsar searches of the Galactic Center to date. No convincing
  candidate was detected in the survey, with a $10\sigma$ detection
  threshold of $\sim 10 \mu$Jy achieved in two separate observing
  sessions.  This survey represents a significant improvement over
  previous searches for pulsars at the Galactic Center and would have
  detected a significant fraction ($\gtrsim 5$\%) of the pulsars around
  Sgr~A*, if they had properties similar to those of the known
  population. Using our best current knowledge of the properties of the
  Galactic pulsar population and the scattering material toward
  Sgr~A*, we estimate an upper limit of 90 normal pulsars in orbit 
  within the central parsec of Sgr~A*.
\end{abstract}

\keywords{Galaxy: center -- pulsars: general}
\section{Introduction}

The detection of radio-emitting neutron stars within the central
parsec of our Galaxy would provide us with an unprecedented
opportunity to study the super-massive black hole \sgra\ and its
environs. For example, a single orbiting pulsar would yield a direct
probe of the magneto-ionized accretion environment around a black
hole, through measurements of temporal changes in the dispersion and
rotation measures \citep{cl97}. Pulsars orbiting within the curved
space-time around \sgra\ (with orbital periods of $\lesssim
100$~years) could serve as probes of gravity in the strong-field
regime, at field strengths far larger than those accessible with
neutron star binaries. The long-term timing of such pulsars,
supplemented by accurate astrometry, would allow precise determination
of their three-dimensional orbital motion around \sgra.  Depending on
the properties of the pulsars and their orbits, it should be possible
to measure subtle general relativistic deviations from Keplerian
orbits (e.g.  time dilation, gravitational redshifts, frame dragging,
Shapiro delays, etc; e.g. \citealp{ckl+04,pl04}), and it may even be
possible to determine the spin of the black hole (e.g.
\citealp{wk99,kbc+04}).

While theoretical estimates indicate that $100 - 1000$ radio pulsars
with periods $\lesssim 100$~years should be orbiting \sgra\ 
\citep{pl04}, the observational evidence for neutron stars at the
Galactic Center (GC) is mostly indirect. For example, recent studies
have found a number of dense clusters of young, massive stars within
$\sim 1$~pc of \sgra\ \citep{sog+03,gsh+05,pgm+06}, while
\citet{wlg06} report X-ray observations of a pulsar wind nebula near
the massive stellar complex IRS~13, with properties consistent with it
being powered by a young neutron star.  Long term monitoring by
\chandra\ has revealed an excess of transient sources within a parsec
of \sgra, interpreted by \citet{mpb+05} as a population of X-ray
binaries. The flaring radio and X-ray source detected by
\citet{bry+05}, $\sim 0.1$~pc from \sgra, is also likely to be an
X-ray binary.

Despite the above evidence for massive stars around \sgra, there is a
remarkable dearth of radio pulsar detections there, despite several
deep searches (e.g.  \citealp{jwv+95,jkl+06,dcl09}). The closest known
radio pulsars are 11\arcmin\ from \sgra, and less than one percent of 
the known pulsar population lies within a degree of the Galactic Center, 
despite indications of a large population in its environs \citep{dcl09}. 
The reason for this deficit is well understood: hyper-strong scattering of radio waves by
the turbulent, ionized gas within the central 100~pc of \sgra, which
results in temporal smearing of pulsed signals. This pulse broadening
has a strong frequency dependence, $\propto \nu^{-4}$, making it
near-impossible to detect pulsars at the typical observing frequencies
of $\lesssim 1.4$~GHz (e.g. \citealp{lc98}).

To overcome the effects of temporal smearing, searches for pulsars at
the GC have been carried out at progressively higher observing
frequencies over the last few years (e.g. \citealp{jkl+06,dcl09}),
albeit as yet without a detection in the central 25~pc. In this work, we report results
from a deep {Green Bank Telescope} (GBT) search for pulsars toward
\sgra\ at $\sim 15$~GHz, the highest observing frequency used till
date. The choice of this frequency is motivated in
Section~\S\ref{sec:strat}, and the observations and results described
in Section~\S\ref{sec:obs}. Finally, Section~\S\ref{sec:dis} discusses
the constraints placed by our observations on the GC pulsar
population, and the prospects for pulsar detections in future surveys.

\section{A Pulsar Search Strategy for the Galactic Center}\label{sec:strat}

Previous surveys of the GC pulsar population have concentrated on
maximizing the likelihood of detection within a few degrees of \sgra\ 
(e.g. \citealp{jkl+06,dcl09}). This has the effect of {\it lowering} 
the optimal observing frequency, because a larger volume can be searched in 
a given amount of integration time at lower frequencies, due to the larger 
telescope beam. However, the region containing pulsars capable of probing 
gravitational effects in the strong-field regime is quite small, $\lesssim 0.4''$, 
much smaller than the GBT beam at even frequencies $\gg 10$~GHz. Since our 
long-term goal is to use pulsars as probes of strong-field gravity around \sgra, 
we optimized our search strategy to detect pulsars in the immediate vicinity of 
(i.e. in orbit around) \sgra\ itself. \citet{cl97} have previously described 
the observational challenges in detecting periodic emission at the GC; their
main arguments are summarized below.

The optimal observing frequency depends on three competing effects,
the temporal scattering, the pulsar emission spectrum, and the
telescope sensitivity, all with very different frequency dependencies.
For example, the generally steep spectral energy dependence of the
pulsar emission ($S_\nu \propto \nu^\alpha$, with the distribution
centered on $\alpha \sim -1.7$) drives searches to low frequencies,
but competes against the strong $\nu^{-4}$ spectral dependence of
temporal smearing, due to which pulsars are severely attenuated at
frequencies below which the scattering time exceeds the pulse period.
We note, in passing, that the intervening ionized plasma also
introduces a dispersive delay in the pulsar arrival times which can
lead to a smearing of the pulse in individual filterbank frequency
channels used to detect the signal, thus degrading pulse
detectability. While this effect has been formally included in the
present analysis, it is insignificant for the fine spectral resolution
offered by modern filterbanks, for the estimated dispersion measure of
$\sim 1500 - 3000$\,pc~cm$^{-3}$ for the GC region \citep{lc98}

The primary uncertainty in the temporal smearing time lies in the
distance of the scattering screen from the Galactic Center. The best
estimate of this distance is from \citet{lc98}, who combined all known
tracers of ionized gas (the scattering diameters of masers and OH/IR
stars, free-free emission and absorption, etc.) in a maximum
likelihood analysis to obtain $D_{\rm scat}=133^{+200}_{-80}\,$pc.
Making the usual simplifying assumption that that the scattering
material is confined to a thin screen results in a temporal smearing
timescale for pulsars near \sgra\ of
\begin{eqnarray}
\tau_{\rm ISM} = 0.116 \left( \frac{D_{\rm scat}}{100\,{\rm pc}}\right)^{-1} 
\left( \frac{\nu}{10\,{\rm GHz}} \right)^{-4}\,\:\:\:{\rm sec}.
\end{eqnarray}

\begin{figure*}
  \includegraphics[angle=270,width=0.9\textwidth]{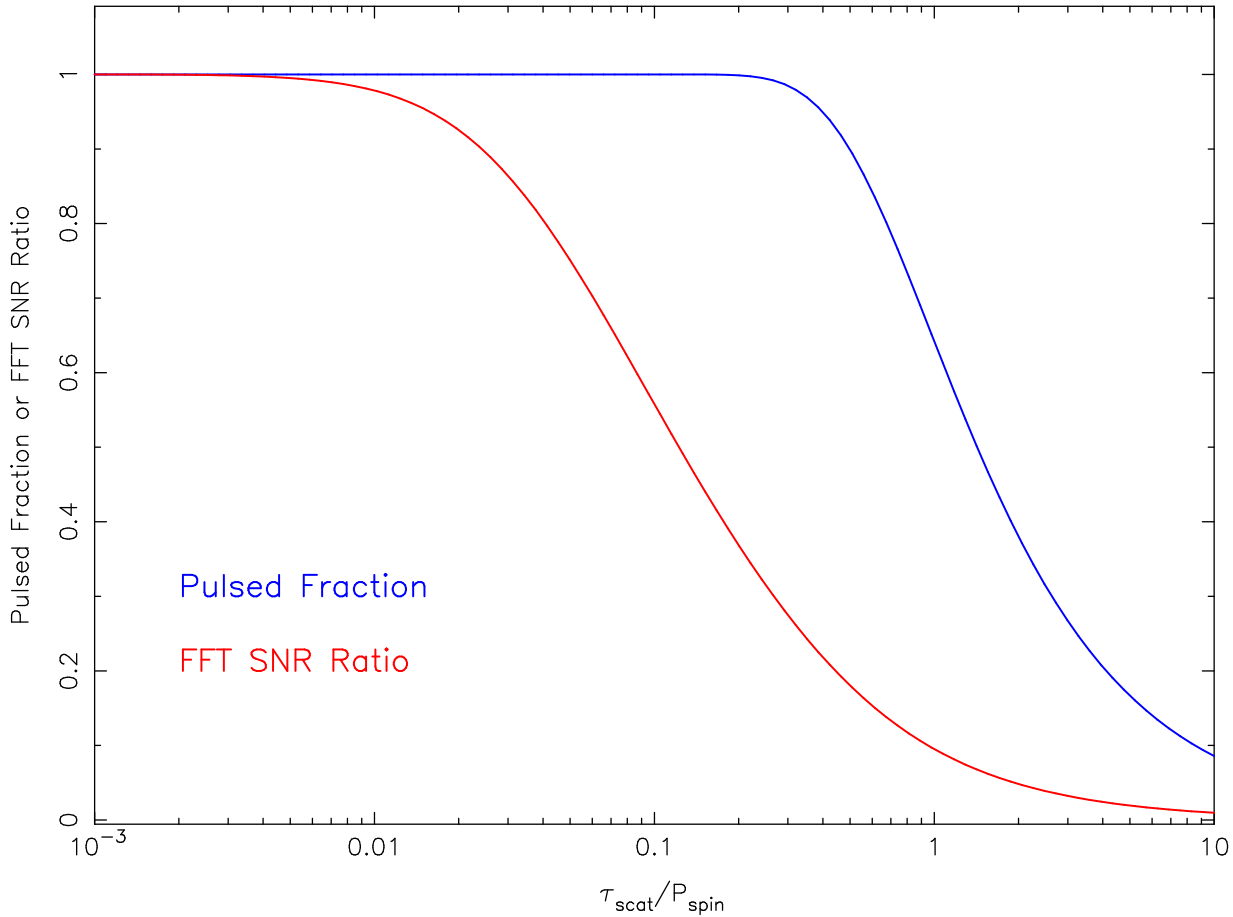}
\caption{The effect of temporal broadening on a pulse with an intrinsic
  pulse width of 5\% of the spin period, $P_{\rm spin}$, assuming a
  one-sided exponential filter for the temporal broadening.  The blue
  line indicates the fraction of the total pulsed flux density from
  the pulsar that remains pulsed after the scattering.  The red line
  depicts the ``Fourier SNR'' (i.e. the sum of average noise-level
  normalized and mean-subtracted Fourier amplitudes) of a 16-harmonic
  summation of the power in a standard Fourier search.  The latter
  indicates the detectability of pulsars in a blind survey and
  implicitly incorporates the pulse profile width increases and pulse
  shape changes caused by scattering.  Most sensitivity is lost beyond
  $\tau_{\rm scat} \sim 0.2\,P_{\rm spin}$.  }
\label{fig:ScatEffectsPlot}
\end{figure*}

At a given frequency, a pulsar with period smaller than $\sim
\tau_{scat}$ would effectively have its pulses smeared into one
another. In other words, for a given observing frequency, it is not
possible to detect a pulsar whose period is much shorter than the 
temporal broadening time at that frequency (see
Fig.~\ref{fig:ScatEffectsPlot}). For instance, a pulsar of period 50\,ms is dominated by
temporal smearing at frequencies $\lesssim 11$\,GHz, while the
detection of a pulsar of period 5\,ms is strongly hampered at
frequencies $\lesssim 26$\,GHz. Combining these effects, we estimate
the temporal smearing time for a pulse of intrinsic width $\tau_{\rm
  intrinsic}$, and with contributions from scattering ($\tau_{\rm
  ISM}$), interstellar dispersion ($\tau_{\rm DM}$) and temporal
binning ($\tau_{\rm res}$) to the overall observed pulse width to be
\begin{eqnarray}
  \tau_{\rm obs} = \left[ \tau_{\rm intrinsic}^2 + \tau_{\rm ISM}^2 +
    \tau_{\rm DM}^2 + \tau_{\rm res}^2 \right]^{1/2}.
\end{eqnarray} 
This equation affords a sufficiently good approximation to the effects 
of temporal broadening, shown in detail in Fig.~\ref{fig:ScatEffectsPlot}, 
when neither the intrinsic pulse profile nor the temporal broadening 
kernel are known in detail.
   
Next, the steepness of the pulsar emission spectrum adversely
impacts searches at high frequencies. The trend from younger to older
pulsars is a steepening of the spectral index, from $-1$ to $-2$
\citep{lylg95} with a mean of $-1.6$. Using a sample of 266~pulsars,
\citet{mkkw00a} find that the flux density typically has a single
power law ($S_\nu \propto \nu^{\alpha}$), with an average spectral
index $\langle\alpha\rangle=-1.8\pm0.2$ \citep{mkkw00a}. For this
paper we use a mean pulsar spectral index of $-1.7$, a value
intermediate between that derived by \citet{lylg95} and
\citet{mkkw00a}. About 10\% of the sample require a dual power law,
with a steeper frequency dependence above $\sim 1.5$\,GHz; these
pulsars would be difficult to detect at high frequencies. Conversely,
there exists a minority population ($<$2\%) of pulsars that show a
flattening or even an upturn in their spectra at higher frequencies
\citep{kxj+96,ljk+08}.

Finally, the frequency dependencies of the performance of the telescope
and its receivers, the Galactic background, and the sky brightness,
all have an impact on the choice of observing frequency. In the case
of the GBT, the receiver contributions to the system temperature are
roughly flat from $1.5 - 20$~GHz. At lower frequencies ($\nu<
10$\,GHz), strong emission from the GC region is a significant (and
sometimes overlooked) contributor to the system temperature, with
T$_{\rm BG}$= $340\,(\nu /2.7\,{\rm GHz})^{-2.7}$\,K \citep{rfrr90}.
Conversely, the atmosphere makes a significant contribution to the
system temperature at high frequencies, $\gtrsim 15$\,GHz, especially
at the low GC elevations when observed with the GBT. The lowest GBT
system temperatures can be shown to arise in the frequency range $\sim
10 - 18$~GHz, with the upper and lower bounds determined by
contributions to T$_{\rm sys}$ from the atmosphere and the GC region,
respectively.

All of the foregoing effects were combined to determine the optimal
observing frequency for the GBT search for pulsars at the Galactic
Center. This was done by computing the signal-to-noise ratio (S/N) for
a 10-hour GBT integration as a function of frequency (using equation~1
of \citealp{dtws85}) for a pulsar with a mean flux density of $S_\nu =
1\,$mJy at a frequency of 1\,GHz, and an intrinsic pulsar duty cycle
of 10\%. This was done for three representative spectral indices,
$\alpha = (-1.0 \;, -1.7\; , -2.5)$ (where $S_\nu \propto
\nu^\alpha$). Finally, three representative periods (5, 50, and
500\,ms) were also used, to illustrate the GBT's sensitivity to
different pulsar populations (corresponding, roughly, to millisecond
pulsars, partially-recycled or young pulsars, and normal pulsars,
respectively).

\begin{figure*}
  \includegraphics[angle=0,width=0.9\textwidth]{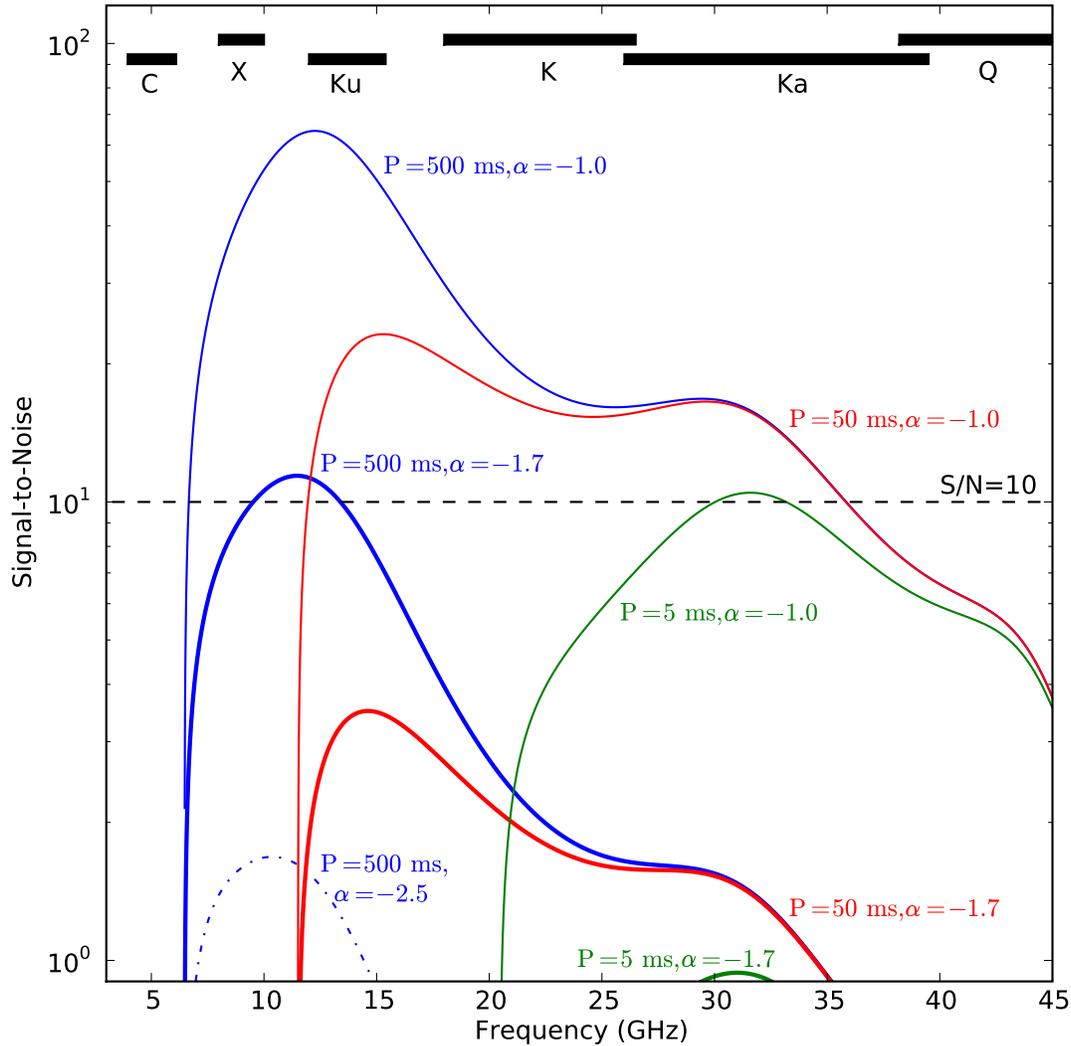}
\caption{The frequency dependence of the detectability of pulsars at
  the Galactic Center for three representative periods, P=5\,ms
  (green), 50\,ms (red) and 500\,ms (blue) for a range of power-law
  spectral indices $\alpha$=-1 (thin line), $-$1.7 (solid line),
  $-2.5$ (dashed line). The vertical axis gives the expected
  signal-to-noise ratio for a $10$\,hour integration with the GBT at
  the Galactic Center, normalized to a pulsar of period 1\,s, flux
  density $S_\nu = 1\,$mJy at a frequency of 1\,GHz, and whose
  spectrum has the frequency dependence $S_\nu \propto \nu^{\alpha}$.
  The frequency range for the current suite of GBT receivers is shown
  at the top of the figure.}
\label{fig:SNplot}
\end{figure*}

The results of this analysis are summarized in Fig.~\ref{fig:SNplot},
where the S/N for a 10-hour GBT integration is plotted versus
frequency; we use S/N\,$= 10$ as the detection threshold. It is clear
from the figure that, for the assumed flux density (1~mJy at 1~GHz),
recycled or millisecond pulsars would not be detectable with the GBT
unless they have flat spectral indices ($\alpha \sim -1$), while
pulsars with $\alpha = -2.5$ would be entirely undetectable.  It is
also clear that the optimal search frequency for ``normal'' pulsars
(period $\sim 500$\,ms, and $\alpha \sim -1.7$) lies in the range
$10-16$~GHz, with the upper end of this frequency range also allowing
the detection of shorter-period, flat-spectrum pulsars. The frequency
range $12-12.8$\,GHz is also affected by strong satellite-based
interference at the GBT. This motivated our choice of 15~GHz as the
optimal observing frequency for the GBT search for pulsars at the
Galactic Center.

\begin{deluxetable}{lcccc}
\tablecaption{GBT Observational Details
\label{table:obs}}
\tablewidth{0pt}
\tablehead{
\colhead{ } & \colhead{$\nu_{obs}$} & \colhead{On-Src} &
\colhead{$T_{sys}$} & \colhead{Sensitivity}\\
\colhead{Date} & \colhead{(GHz)} & \colhead{Time (hrs)} &
\colhead{(K)\tablenotemark{a}}& \colhead{($\mu$Jy)\tablenotemark{b}}
}
\startdata
22 Jun 2006 & 14.8 & 2.5 & 32 & 18  \\ 
28 Jun 2006 & 14.8 & 4.5 & 38 & 16  \\ 
29 Jun 2006 & 14.8 & 5.5 & 32 & 12  \\ 
10 Aug 2006 & 14.8 & 3.0 & 37 & 19  \\ 
11 Aug 2006 & 14.8 & 2.25 & 50 & 30 \\ 
30 Aug 2008 & 14.4 & 4.25 & 36 & 16 \\ 
31 Aug 2008 & 14.4 & 5.5  & 33 & 13 \\
\enddata
\tablenotetext{a}{The quoted ${\rm T_{sys}}$ values are the averages
for each run, after correcting for the different elevations of the GC
and the calibrators.}
\tablenotetext{b}{The last column lists the $10\sigma$ detection
  threshold flux density for pulsars with a duty cycle of 10\%, in
  each individual observing session. Note that the final searches
  combined data from adjacent days, and hence had a lower detection
  threshold flux density, $\sim 10 \mu$Jy apiece for the runs of 28
  and 29~June, 2006, and 30 and 31 August, 2008.}
\end{deluxetable}

\section{Observations, data analysis, and results}
\label{sec:obs}

\subsection{The observations}
\label{subsec:obs}

The search for pulsars at the Galactic Center was carried out with the
Ku-band receiver of the GBT in the summers of 2006 and 2008, with the
GBT Pulsar Spigot \citep{kel+05} used as the backend. The observations
were typically restricted to GC elevations higher than 10~degrees, to
reduce atmospheric contributions to the system temperature (${\rm
  T_{sys}}$).  The GC is only visible with the GBT for $\sim 6$~hours
above this elevation limit per day, and the observations were
therefore broken up into multiple short sessions. While attempts were
made to group these sessions as close together as possible (so as to
combine multiple sessions in a search), this was often not possible
due to poor high-frequency observing weather.

The 2006 observing runs were on June~22, 28 and 29, July~10, and
August~10 and 11.  The July~10 data were affected by strong broad-band
radio frequency interference (RFI) due to an overloaded network
switch, and will not be discussed further; data from the other runs
were not affected by this issue. The 2008 observations were on
August~30 and 31, with each run preceded by a 5-hour observation of a
blank field (J2000 co-ordinates: Right Ascension=11h46m08.1s,
Declination=$-$27d47'32.9''), with the same observing parameters as
for the GC; the latter served to test for systematic effects in the
data.  The on-source times for the GC are listed in column~(3) of
Table~\ref{table:obs}.

All observing runs included short observations of strong known pulsars (B1800$-$21 in 
most of the 2006 runs and the magnetar, XTE~J1810$-$197, in August~2006 and 2008), 
to test that the system was working properly; these were clearly detected in every run.

All observing sessions used the maximum allowed Spigot bandwidth of
800~MHz, sub-divided into 1024~channels, with two circular
polarizations, 3-level sampling, and a dump rate of 81.92$\:\mu$s. The
800~MHz band was centered at 14.8~GHz in the 2006 runs, and at
14.4~GHz in the 2008~runs; the shift in the central frequency was to
exclude some RFI (at 14.88~GHz) from the observing band. In 2006, an
automatic level controller was used to hold the power levels fixed on
the GC.  Unfortunately, this was found to itself result in
intermittent power jumps, and we hence chose to not use this feature
in 2008. In all runs, observations of strong nearby calibrators (or
the GC itself) were used to correct the telescope pointing and
surface, typically every $2-3$~hours. System temperatures were
measured on the calibrators by firing a noise diode, and ranged
between 26~K and 40~K on most observing runs; however, the short
session on August~11 had ${\rm T_{sys}} \sim 50$~K on the calibrator,
due to overcast conditions.  After taking into account the slightly
different elevations of the GC and the calibrators, the estimated
average system temperatures towards the GC were $32 - 38$~K (again,
except on 11~August, where ${\rm T_{sys}} \sim 50$~K).  The gain of
the Ku-band receiver is 1.5 K\,Jy$^{-1}$.  Finally, the size of the
GBT beam at 14.8~GHz is $\sim 50''$, corresponding to a spatial radius
of $\sim 1$~pc at the distance of the GC ($7.9^{+0.8}_{-0.7}$~kpc;
\citealp{rmz+09}).

Table~\ref{table:obs} presents a summary of the observational details;
the columns are (1)~the observing date, (2)~the central frequency,
(3)~the GC on-source time, (4)~the typical system temperature on the
GC during the run, and (5)~the $10\sigma$ detection threshold for
pulsars with a duty cycle of 10\% in each session (using equation~1 of
\citealp{dtws85}).  As discussed below, the final searches were
carried out on the combined datasets from (1)~June 28 and 29, 2006
($\sim 10$~hours), (2)~August 10 and 11, 2006 ($\sim 5.25$~hours), and
(3)~August 30 and 31, 2008 ($\sim 9.75$~hours). The first and third of
these had the best sensitivity, with $10\sigma$ detection thresholds
of $\sim 10 \mu$Jy each, again assuming a pulsar duty cycle of $10$\%.

\subsection{Data analysis}
\label{subsec:data}

The observations were reduced and analyzed with the 
PRESTO\footnote{See {\tt http://www.cv.nrao.edu/$\sim$sransom/presto/}.}
software package \citep{ran01}.  Initially, all data were time-domain clipped 
at the $6\sigma$ level to remove any strong RFI.
As noted above, the 2006 data suffered from random jumps in the power
level due to an automatic level controller in the signal path.  The
effect of these jumps was removed by forcing data between jumps to a
common mean of zero, with data in the immediate vicinity of the jump
points edited out. Data from successive days (28 and 29~June 2006, 10
and 11~August 2006, and 30 and 31~Aug 2008) were combined into a
single data stream in order to enhance the detectability of faint
pulsars during the periodicity search. The data from 28 and 29~June 2006 
and 10 and 11~August 2006 were also analysed independently, to examine the
possibility of intermittent signals. The data streams were de-dispersed 
for dispersion measures of 1500, 2000, 2500, 3000, 3500,
4000 and 5000~cm$^{-3}$\,pc.  Finer sampling in dispersion measure is
not required for the detection of slow pulsars because the relative
dispersion delay across the 800\,MHz band at 14.8\,GHz is only 2\,ms
for every 1000\,cm$^{-3}$\,pc.  Note that our search was only
sensitive to pulsars with periods $< 2$\,s, due to power fluctuations
associated with atmospheric variability that reduced our sensitivity
and precluded a search for pulsed emission at longer periods.

The search for pulsars was carried out using standard techniques: the signal 
was Fourier transformed, a red-noise reduction was applied to the spectrum,
and the PRESTO algorithm {\it accelsearch} \citep{rem02} used to
identify potential pulsars. The original time series were then folded
according to the pulse characteristics of each candidate in order to
determine the significance of each signal, construct its pulse
profile, and refine its characteristics (e.g. period, period derivative, 
etc.)

We also separately searched the 2008 data for bright individual pulses, like those 
seen from young pulsars such as the Crab, using the program {\tt single\_pulse.py} 
from PRESTO, a time-domain matched-filtering technique similar to that
described by \citet{cm03}. The de-dispersed time series were first down-sampled 
by a factor of 10 to an effective time resolution of 0.8192\,ms.  We then 
convolved the data with square-wave pulses of a variety of widths from 1 
to 150 samples in duration and searched for peaks substantially above those of 
the noise variations.  In no case did we find strong pulses that were obviously 
dispersed in nature and could be clearly distinguished from RFI. Of course, a 
significantly larger observing bandwidth would provide not only more sensitivity 
to pulses of this sort, but would also dramatically increase the amount of 
dispersive smearing across the band, allowing us to use dispersion to rule out 
pulses of terrestrial origin.  While we also searched the 2006 data for single-pulse 
signals without any detections, the sharp jumps introduced by the auto-leveler in these
datasets rendered it particularly difficult to distinguish astronomical pulses
from system-generated signals.

\subsection{Results}\label{subsec:results}

\begin{figure}[htp]
\begin{center}
\includegraphics[angle=0,width=0.48\textwidth]{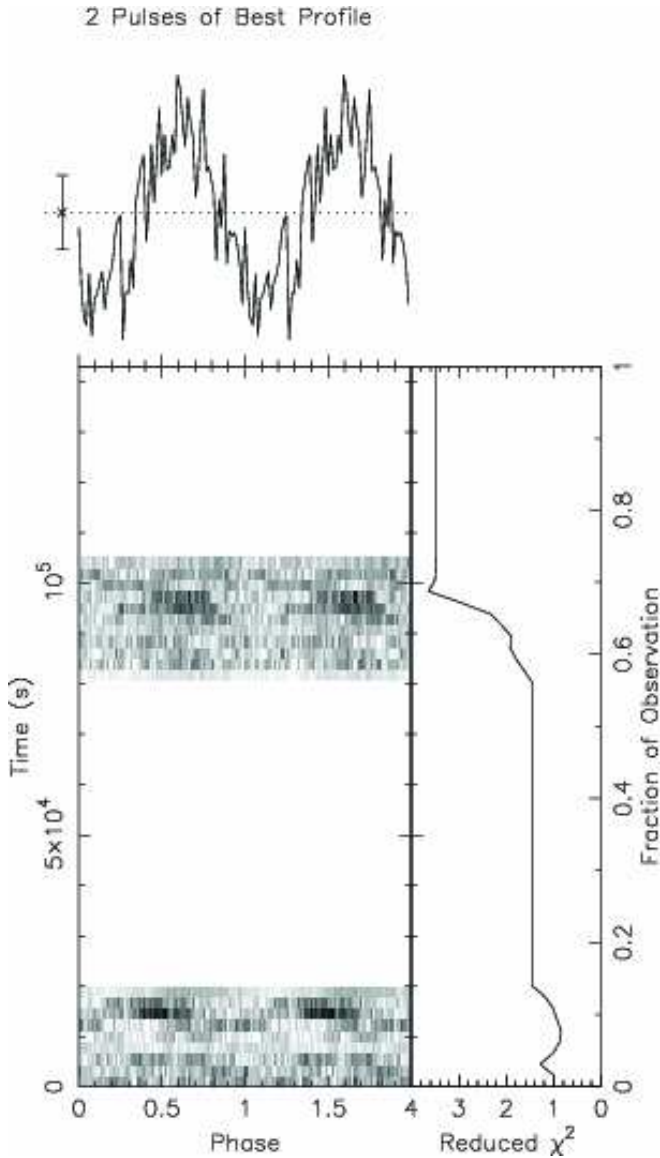}
\caption{A detection plot of the putative 607\,ms pulsar candidate from the dataset 
of 28-29 June 2006; the top panel shows the final pulse profile, while the bottom 
panel shows the growth of reduced $\chi^2$ with increasing data.  The grayscale denotes the signal strength, with darker signals indicating a stronger signal. The white regions correspond to time intervals during which the telescope was not pointing at the Galactic Center.} \label{fig:Res1}
\end{center} 
\end{figure}

\begin{figure}[htp]
\begin{center}
\includegraphics[angle=0,width=0.48\textwidth]{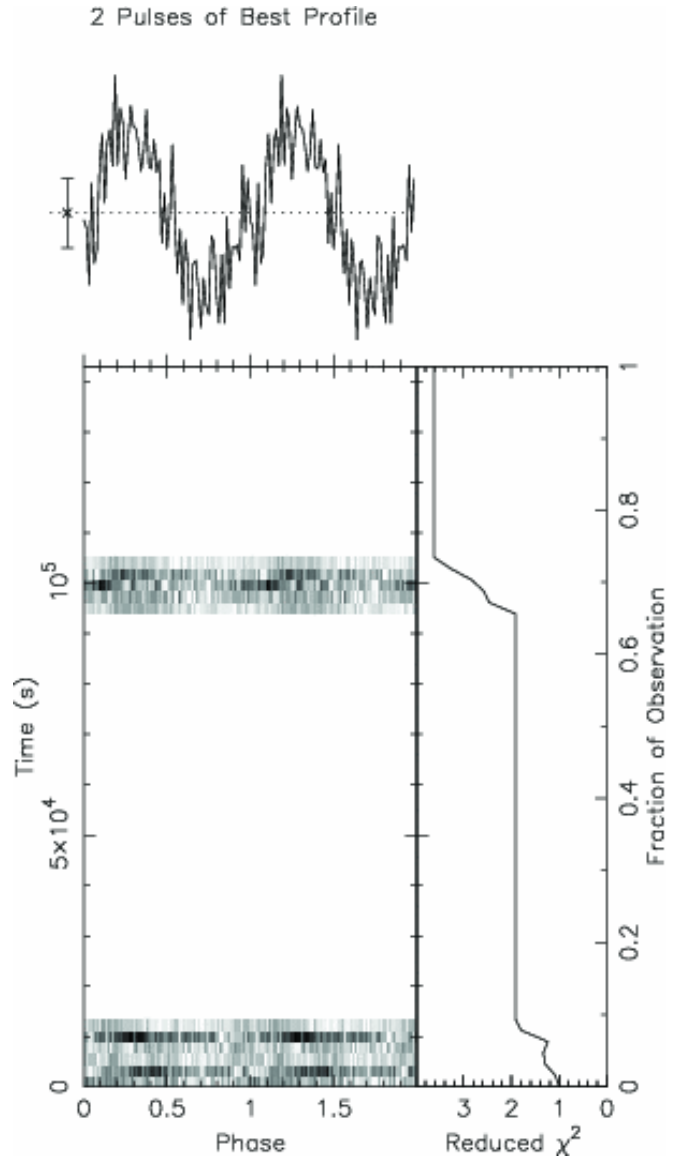}
\caption{Left: A detection plot of the putative 604\,ms candidate from the dataset of 
10-11 August 2006.  } 
\label{fig:Res2}
\end{center} 
\end{figure}

\begin{figure}[htp]
\begin{center}
\includegraphics[angle=0,width=0.48\textwidth]{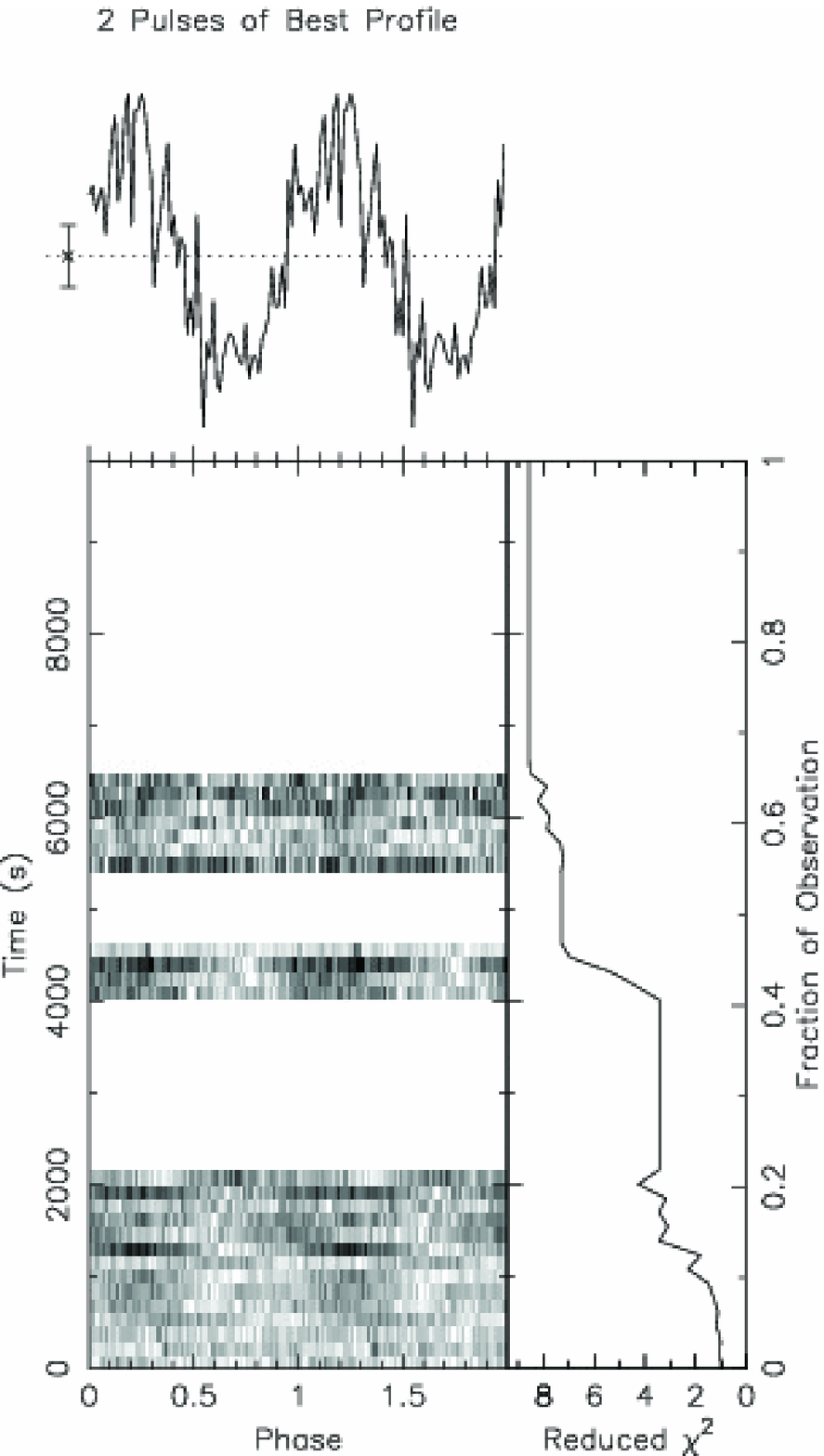}
\caption{A detection plot of the pulsed signal on 22 June 2006, after folding at
the pulse period of the candidate detected in the dataset of 28-29 June 2006.} \label{fig:Res3}
\end{center} 
\end{figure}

Our original search in 2006 yielded candidate pulsed signals of high
single-trial statistical significance ($> 8\sigma$) in the 28-29~June and
10-11~August datasets, independently. The highest S/N was obtained at
a trial dispersion measure of $3000$\,pc\,cm$^{-3}$, and a period of
607\,ms in the dataset of 28-29 Jun 2006 (11$\sigma$ significance);
this pulse profile is shown in Fig.\,\ref{fig:Res1}.  Signals at the
harmonic periods of 303\,ms and 152\,ms were also weakly detected in
this dataset, typically at $\sim 4\sigma$ significance. A similar
signal, with a period of $\sim 604$\,ms, was independently detected in
the dataset of 10-11 August, with a significance of $\sim 8\sigma$ at
trial dispersion measures of $3000 -4000$\,pc\,cm$^{-3}$; this is
shown in Fig.\,\ref{fig:Res2}). Again, harmonics of this signal (at
302\,ms and 151\,ms) were detected at lower significance ($\sim
3.5\sigma$).
The 607\,ms signal was also seen on folding the data from 22~June 2006 data using the 
solutions determined from the dataset of 28-29 June; this yielded a detection significance 
of $18 \sigma$, and the pulse profile of Fig.\,\ref{fig:Res3}. We also tested that 
these candidate pulsed signals do not arise from periodicities in the power jumps in 
the 2006 data.

No evidence for a pulsed signal was found in the dataset from 2008 at these, or any 
other, periods. The 2008 data were also folded at the pulse periods of the candidates 
detected in 2006 ($\sim 604-607$\,ms), without any detectable signal. 

Finally, no other pulsar candidates (above 6$\sigma$ significance) were found in any
of the datasets.  The noise levels achieved in each observation are listed in Table 
\ref{table:obs}. The best sensitivity was achieved with the datasets of 28-29~June 2006, 
and 30-31~August 2008, which yielded a $10\sigma$ detection threshold of $10\mu$Jy, 
assuming a pulsar duty cycle of 10\%.

\section{Discussion and Conclusions} \label{sec:dis}

\subsection{The 607\,ms candidate of 2006}\label{sec:candidate}

The high statistical significance of the pulsed signal detected in the 
datasets of 2006 implies that it arises either from a genuine pulsar 
towards the GC or as an artifact of a terrestrial signal (e.g. RFI). While 
the non-detection in the 2008 dataset might indicate the latter possibility, 
it should be emphasized that the GC environment is very different from 
the environments of typical pulsars. Specifically, a pulsar on a 
short-period ($< 100$~year) orbit around the GC could easily have its 
emission beam precess away from our line of sight over a time-scale of 
2~years. This implies that caution must be used while dismissing 
possible pulsar candidates towards the GC, although retaining skepticism 
about their reality. We will hence summarize the characteristics of 
the pulsed signals seen in the 2006 datasets, and discuss the 
possibilities that they might arise from a real pulsar or RFI.

Figs.\,\ref{fig:Res1} and \ref{fig:Res2} show that the detection $\chi^2$ 
in the two long datasets of 2006 increases steadily over the course of 
each observing session, indicating that, if the signals are spurious or 
local RFI, they are at least persistent both over the course of each 
observation, and over multiple observing epochs. It is curious, however, 
that the period of the pulsed signal is different in the sessions on 
28-29~June and 10-11~August; for RFI, this would require either that we have detected 
two distinct but alternately intermittent RFI signals, or that the period 
of the RFI itself is changing. Interpreted in terms of Doppler shifts, 
the period change corresponds to a velocity change of 1500~km/s, much 
larger than that associated with the Earth's motion around the Sun or 
the motions of terrestrial objects, but not implausible for a pulsar 
orbiting around the GC.

Next, the S/N ratios of our candidates peak at dispersion measures of 
$\approx 3000-4000$\,pc~cm$^{-3}$, comparable to values expected for 
pulsars at the GC.  Unfortunately, our small fractional bandwidth means 
that the dispersion in the signal across the band is very small 
($\sim 3.6$\,ms over the 800~MHz bandwidth). This means that, unlike 
the situation in low-frequency pulsar surveys, dispersion cannot be used 
to test whether the signal is of extra-terrestrial origin. Finally, the 
pulse profiles of the candidates are extremely broad, with a duty cycle 
of $\sim 50$\%, unlike the narrow profiles expected for high-frequency 
pulsar emission. However, the pulse properties too could be affected 
by the unique GC environment. For example, the thin-screen approximation 
might not be applicable for the scattering, or the screen could be 
much closer to the pulsar than typical estimates of $\sim 100$~pc; both 
of these would increase the scattering time and broaden the pulse profile,
even at such a high frequency. Specifically, the scattering timescale at a 
frequency of 14.6~GHz for an object at the Galactic Center is 
$2.5/D_{\rm scat}$~seconds, where $D_{\rm scat}$ is the distance, in pc, of 
the scattering medium from \sgra. While the best estimate of $D_{\rm scat}$ is 
$\sim 100$~pc from angular broadening measurements of \sgra\ and nearby masers 
\citep{cl97}, the effect of scattering material close to \sgra\ is much stronger 
on temporal smearing than on the angular broadening of background sources. As such,
the angular broadening estimates of $D_{\rm scat}$ do not rule out a substantial 
contribution to the pulse broadening from material closer to \sgra. One may hence 
have a sizeable contribution to the pulse broadening from material at $D_{\rm scat} 
\lesssim 10$~pc (e.g. \citealt{mb06}). The expected temporal smearing timescale would
then be $\gtrsim 250$~ms, comparable to that needed to explain the pulse shape 
of the 607~ms candidate. The large observed duty cycle of the candidate thus does 
not rule out the possibility that the signal arises from a genuine pulsar.

It thus appears very difficult to rule out the reality of the candidate 
on the basis of the 2006 data alone, and, as noted above, the non-detection 
in 2008 could arise due to precession of the pulsar beam away from 
our sightline. Thus, while we remain skeptical about the reality of 
these signals, we conclude that further observations are needed to test 
the possibility that they arise from a genuine pulsar at the GC. 

\subsection{Constraints on the GC pulsar population}\label{sec:constraints}

There is compelling but indirect evidence for a substantial population
of neutron stars at the Galactic Center. However, strong interstellar
scattering along the line of sight has limited past searches for radio
pulsars. To overcome these effects, we have used the superb
sensitivity of the GBT to carry out a deep search for pulsars in the
central parsec of the GC at 15~GHz --- the highest observing frequency
at which a search has been carried out to date.  Despite this, we find
no convincing pulsar candidates. Was our survey sufficiently sensitive
to detect a population of pulsars around \sgra?

The total number of pulsars detectable at the GC depends on the total
number of pulsars accumulated in the region, and the fraction of these
objects that would be detectable given our survey sensitivity, and the
S/N considerations of Section~\ref{sec:strat}. The detectable fraction
depends particularly on the number of pulsars with flat spectral
indices, since these objects influence the pulsar luminosity function
most strongly at frequencies $> 10$\,GHz where they are most easily
detectable towards \sgra.

\begin{figure*}
\includegraphics[angle=0,width=0.9\textwidth]{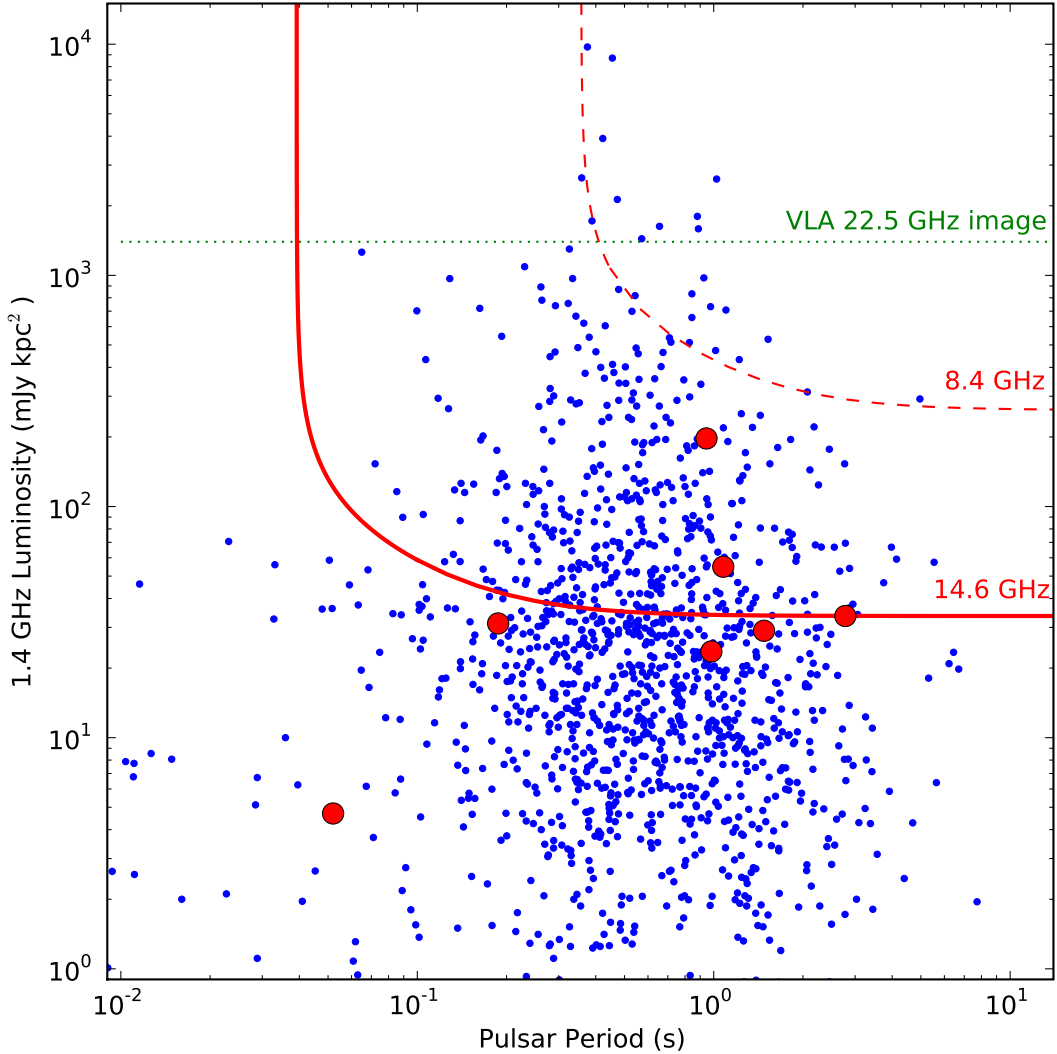}
\caption{The 1.4 GHz luminosities of the known sample of pulsars 
  versus pulse periods (blue dots). Larger circles (red dots) indicate
  those pulsars within one degree radius of \sgra. The $10\sigma$ pulsar
  sensitivity of our 14.6\,GHz search is shown by the solid red line.
  This was obtained by using the flux density limit (10 $\mu$Jy at 10$\sigma$) 
  of our survey to calculate the luminosity limit at the distance of the
  Galactic Center and then scaling the result to a frequency of 1.4\,GHz, 
  using an average spectral index of $\langle\alpha\rangle=-1.7$. We also show 
  sensitivity curves derived in the same manner for a Parkes 8.4\,GHz pulsar 
  survey of the GC (the dashed red line, with a $10\sigma$ detection threshold of 
  200\,$\mu$Jy; \citealp{jkl+06}), and a deep 22.5\,GHz VLA image of the GC 
  (the dotted green line, with a $10\sigma$ detection threshold of 200\,$\mu$Jy; 
  \citealp{zmga09}).}
\label{fig:sens}
\end{figure*}

A simple estimate of the number of detectable pulsars can be obtained
by positing that the \sgra\ pulsar population has similar properties
to those of the {\it known} population of pulsars and to estimate the
fraction of the known population that would be detectable at the GC
with our survey. This is done in Fig.\,\ref{fig:sens}, where we have
plotted pulsars with measured 1.4\,GHz luminosities [from the
\citet{mhth05} catalog] on a period-luminosity diagram. The solid red
line shows the pulsar sensitivity curve of our 14.6\,GHz survey,
obtained using equations\,(1-3) with a 10$\sigma$ detection threshold
of 10 $\mu$Jy, and assuming a 10\% pulsar duty cycle, a scattering
screen distance $D_{\rm scat}=133$\,pc (\S\ref{sec:strat}) and a GC
dispersion measure of $1700$\,pc~cm$^{-3}$.  The sensitivity curve has
been scaled to 1.4\,GHz using a mean pulsar spectral index of $-1.7$.
The cutoff in period where most of the sensitivity is lost is taken to
be at P$_{\rm spin}$=$2\times\tau_{\rm scat}$. This is less severe
than the scatter-based sensitivity cutoff in
Fig.~\ref{fig:ScatEffectsPlot} but does reflect the fact that some
partially recycled or young pulsars (i.e. P$<50$ msec) would be
detectable if they were much brighter than our noise threshold.

For comparison purposes, this figure also shows the 8.4\,GHz
sensitivity curve for the Parkes GC survey \citep{jkl+06}, and the
$5\sigma$ noise threshold for a deep imaging survey of the GC at
22.5\,GHz, using the Very Large Array \citep{zmga09}, again scaling
both of these to a frequency of 1.4\,GHz using a mean spectral index
of $-1.7$.  We also highlight the seven known pulsars within one
degree of \sgra, including four new ones from \citet{dcl09} and
\citet{cng+09}.

A more rigorous estimate can be obtained by computing the fraction of
pulsars detectable above some flux density cutoff by considering the
pulsar luminosity function at $\nu_0=1.4$\,GHz, $f_0 (L)$, combined
with the spectral index distribution, $p(\alpha)$. This is the
approach followed by \citet{pl04} and \citet{cl97}, but updated with
the most recent results on pulsar luminosity functions
\citep{fk06,lfl+06}. We model the 1.4\,GHz luminosity function as a
power law between lower and upper cutoffs $L_{\rm min}$ and $L_{\rm
  max}$ respectively:
\begin{eqnarray}
f_0(L) = A L^{-\beta},  \qquad A=(1-\beta) \left[ L_{\rm max}^{1-\beta} - 
L_{\rm min}^{1-\beta} \right]^{-1}  \;\;,
\end{eqnarray}
where the normalization is chosen so that the integral over all
luminosities is unity, such that $f dL$ is interpreted as the fraction
of all pulsars with luminosities between $L$ and $L+dL$. 
Recent studies suggest that $L_{\rm min}=0.01\,$mJy\,kpc$^2$, 
$L_{\rm max}=32\,$Jy\,kpc$^2$ and $\beta = 1.2-2$ \citep{fk06,lfl+06}. 
Following \citet{sks+09}, the spectral index distribution is modeled as a 
Gaussian
\begin{eqnarray}
p(\alpha)= \frac{1}{\sqrt{2 \pi \sigma_\alpha^2}} \exp \left[ - 
\frac{(\alpha-\alpha_m)^2}{2 \sigma_\alpha^2}\right] \;\;,
\end{eqnarray}
with mean spectral index $\alpha = -1.7$ and standard deviation $\sigma_\alpha =0.35$.

\begin{figure*}[t]
\begin{center}
  \includegraphics[angle=0,width=0.9\textwidth]{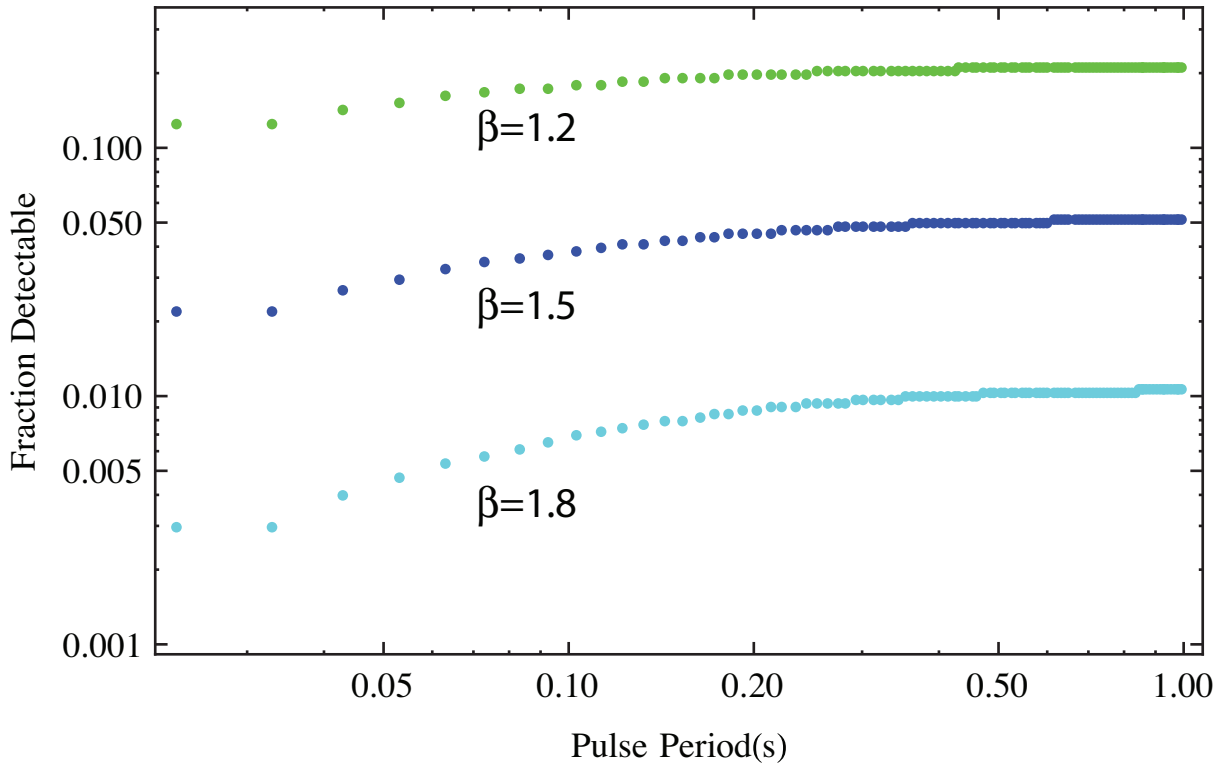}
\caption{The fraction of Galactic Center pulsars that would be
  detectable at $10\sigma$ significance for a 10\,h integration on the
  GBT at 15~GHz as a function of spin period. We assume an intrinsic
  pulse width of 10\% of the spin period, and take $D_{\rm
    scat}=100\,$pc.  The curves, from top to bottom, correspond to
  luminosity function indices of $\beta=1.2, 1.5$ and $1.8$, with a
  mean spectral index of $\alpha = -1.6$ and dispersion
  $\sigma_\alpha=0.35$, as discussed in the text.} \label{FracDetect2}
\end{center} 
\end{figure*}

The pulsar luminosity function at some arbitrary frequency is then
\begin{eqnarray}
  f(\nu,L ) = \int_{-\infty}^{\infty} d \alpha \, p(\alpha) f \left( L
    \left( \frac{\nu}{\nu_0} \right)^{-\alpha} \right),
  \label{lumfrac}
\end{eqnarray}
where, for a given spectral index $\alpha$ chosen from the distribution, the luminosity function
$f$ has lower and upper cutoffs
\begin{eqnarray}
L_{\rm min}'  = L_{\rm min} \left( \frac{\nu}{\nu_0} \right)^{\alpha}, \qquad L_{\rm max}' =  L_{\rm max} \left( \frac{\nu}{\nu_0} \right)^{\alpha}, 
\end{eqnarray}
and the normalization constant $A$ is modified to
\begin{eqnarray}
A = (1 - \beta) \left( \frac{\nu}{\nu_0} \right)^{-\alpha \beta} \left( L_{\rm max}'^{1 - \beta} - L_{\rm min}'^{1-\beta} \right)^{-1}
\end{eqnarray}
in order to ensure that $f dL$ may be interpreted as the fraction of all pulsars in 
the luminosity range $L$ to $L+dL$. We integrate 
equation\,(\ref{lumfrac}) to obtain the total fraction of pulsars above a given 
flux density threshold $L_{\rm cut} = d^2 S_{\rm cut}$.   Of course, this threshold 
depends on both the pulsar period and observing frequency due to propagation effects and 
changes in the system temperature. Figure \ref{FracDetect2} hence plots the fraction of 
the pulsar population detectable above $10\sigma$ significance at 15\,GHz versus 
pulse period.

Note that the present survey is sensitive only to the slower
(P$>40$\,ms) and more-luminous (L$>$40\,mJy\,kpc$^2$) pulsars. The
millisecond pulsars that are presumably powering the low-mass X-ray
binaries near \sgra\ \citep{mpb+05} and the low-luminosity tail of
young pulsars \citep{cng+09} are out of the reach of our survey.
However, it is also clear from the figures that this is the first
survey capable of peering past the ``fog'' of scattering material and
detecting a significant number of pulsars within a parsec of the GC
with properties similar to the known pulsar population. Past
high-frequency pulsar searches \citep{jkl+06} or imaging searches
\citep{zmga09} have not had the requisite temporal or flux density
sensitivity to detect a significant fraction of the known population.
The GBT search thus represents a significant improvement over past
pulse searches and imaging efforts.  We note, in passing, that this
implicitly assumes that pulsar spectral indices do not typically
steepen at high frequencies, $\gtrsim 5$~GHz.

It is clear from Figs.\,\ref{fig:sens}-\ref{FracDetect2} that the
15\,GHz GBT survey could have detected a significant fraction
($\sim$ 1\%-15\%) of the pulsars around \sgra, if they had properties
similar to those of the known population. The estimate obtained from
Fig.\,\ref{fig:sens} is at the high end (15\%) and it is likely biased
by luminosity-dependent completeness limits in pulsar surveys. The
lowest estimate ($\sim$1\%) comes from the curve in
Figs.\,\ref{FracDetect2} with the steepest luminosity slope
($\beta=1.8$). We adopt a nominal value of 5\% from the intermediate
curve ($\beta=1.5$) which is based on our best current knowledge of
the properties of the pulsar population and the scattering material
toward \sgra. Given this detection fraction and our null detection we
can use straightforward binomial statistics to estimate the size of
the putative pulsar population at the GC. If the probability of
detecting a normal pulsar is 5\%, the non-detection of any pulsars in 
our survey implies the upper limit (at 99\% confidence level) of 90~normal 
pulsars within the 1~pc region around the GC encompassed by the GBT beam. 
Taking the full allowed range of the detection fraction (1\% to 15\%), the 
upper limit on the number of normal pulsars ranges from 460 to 30, respectively,
again at 99\% confidence level.  Although our estimate is both approximate and 
subject to much uncertainty, we note that it is significantly lower than the 
$\sim 100-1000$ pulsars derived by \citep{pl04} for the normal pulsar population 
with orbits of $\leq$100 yrs (i.e. a radius 50 times smaller than the size 
of the GBT search area).

Finally, we have shown that the frequency range $10-16$\,GHz is
optimal for searches for ``normal'' pulsars at the GC. The GBT remains
the most powerful high-frequency instrument capable of detecting the
GC pulsar population for at least the next decade, until the advent of
next-generation telescopes like the Square Kilometer Array. The
primary limitation of the GBT (and of the present survey) is the
relatively small instantaneous bandwidth (800~MHz) available for such
searches, resulting in a small fractional bandwidth. An increased
bandwidth at the GBT would imply not only an improvement in
sensitivity, but also a better rejection of terrestrial signals, using
the dispersive sweep of genuine signals across the band.
Fig.\,\ref{fig:sens} shows that an improvement in sensitivity by
merely a factor of $2-3$ would push the GBT into the bulk of the
pulsar population. Future GBT experiments should hence aim to utilize
the full frequency coverage available with the high-frequency
receivers, for both better discrimination against systematic effects
and improved sensitivity.

\acknowledgements

The Green Bank Telescope is operated by the National Radio Astronomy
Observatory, a facility of the National Science Foundation operated
under cooperative agreement by Associated Universities, Inc. We thank
Carl Bignell for his extraordinary efforts to schedule this project on
the GBT. NK is grateful for support from the Max-Planck Society and
the Alexander von Humboldt Foundation, as well as from a Ramanujan Fellowship. 
NK and JPM also acknowledge support from NRAO Jansky Fellowships.

{\it Facilities:} \facility{GBT}

\bibliography{journals_apj,psrrefs,otherpsr}

\clearpage

\end{document}